\documentclass[english]{article}
\usepackage[T1]{fontenc}
\usepackage[latin9]{inputenc}
\usepackage{booktabs}
\usepackage{units}
\usepackage{amsmath}
\usepackage{amsthm}
\usepackage{amssymb}

\makeatletter

\providecommand{\tabularnewline}{\\}

\numberwithin{equation}{section}
\numberwithin{figure}{section}
\theoremstyle{plain}
\newtheorem{thm}{\protect\theoremname}
\theoremstyle{remark}
\newtheorem{rem}[thm]{\protect\remarkname}

\usepackage{babel}
\usepackage{fullpage}
\usepackage{authblk}
\usepackage{hyperref}
\usepackage{placeins}
\usepackage{hyphenat} 
\usepackage{graphicx}
\usepackage{fancyvrb}
\usepackage[hypcap]{caption}
\usepackage[title,titletoc]{appendix}
\addto\captionsenglish{}

\title{Computing XVA for American basket derivatives \\ by Machine Learning techniques}
\author{ \textsc{Ludovic Goudenege}\thanks{F\'ederation de Math\'ematiques de CentraleSupélec - CNRS FR3487, France -\texttt{ ludovic.goudenege@math.cnrs.fr}} 
\and \textsc{Andrea Molent}   \thanks{Dipartimento di Scienze Economiche e Statistiche, Universit\`a degli Studi di Udine, Italy - \texttt{andrea.molent@uniud.it}} 
\and \textsc{Antonino Zanette}\thanks{Dipartimento di Scienze Economiche e Statistiche, Universit\`a degli Studi di Udine, Italy - \texttt{antonino.zanette@uniud.it}}}
\date{}

\def\R{{\mathbb R}}

\numberwithin{equation}{section}
\numberwithin{figure}{section}

\newtheorem{Prop}{Proposition}[section] 

\makeatother

\usepackage{babel}
\providecommand{\remarkname}{Remark}
\providecommand{\theoremname}{Theorem}

\begin{document}
\maketitle

\begin{flushleft}
\rule{1\columnwidth}{1pt}
\par\end{flushleft}

\begin{flushleft}
\textbf{\large{}Abstract}{\large\par}
\par\end{flushleft}

\noindent Total value adjustment (XVA) is the change in value to
be added to the price of a derivative to account for the bilateral
default risk and the funding costs. In this paper, we compute such
a premium for American basket derivatives whose payoff depends on
multiple underlyings. In particular, in our model, those underlyings
are supposed to follow the multidimensional Black-Scholes stochastic
model. In order to determine the XVA, we follow the approach introduced
by Burgard and Kjaer \cite{burgard2010pde} and afterward applied
by Arregui et al. \cite{arregui2017pde,arregui2019monte} for the
one-dimensional American derivatives. The evaluation of the XVA for
basket derivatives is particularly challenging as the presence of
several underlings leads to a high-dimensional control problem. We
tackle such an obstacle by resorting to Gaussian Process Regression,
a machine learning technique that allows one to address the curse
of dimensionality effectively. Moreover, the use of numerical techniques,
such as control variates, turns out to be a powerful tool to improve
the accuracy of the proposed methods. The paper includes the results
of several numerical experiments that confirm the goodness of the
proposed methodologies. 

\vspace{2mm}

\noindent \emph{\large{}Keywords}: XVA; Gaussian Process Regression;
Basket option; Control variates

\noindent \rule{1\columnwidth}{1pt}

\newpage

\section{Introduction}

After the financial crisis of 2007 and the default of several financial
institutions, practitioners, regulators, and finally, academics have
turned increasing attention to counterparty risk. Currently, careful
and weighted management of counterparty risk is required at the legislative
level by the Basel III agreements of 2010, as well as codified by
the IFRS standard starting from 2013. Consequently, when assessing
the value of an OTC derivative instrument, banks must apply a series
of adjustments to the risk-free price, capable of accounting for the
costs associated with the effects of a possible default of any of
the two counterparties. The entirety of these corrections is known
as the total credit value adjustment, usually indicated by the abbreviation
XVA. The main elements that contribute to the calculation of the XVA
are the CVA, the DVA and the FVA. The CVA, credit value adjustment,
is the premium that an agent must charge to the counterparty to cover
the losses that could derive from the default of the same. In particular,
these losses occur when the value of the contract is positive for
the agent and the counterparty, following a default, is unable to
comply with the contractual terms. The DVA, debit value adjustment,
is the consideration of the CVA for the counterparty: in the event
of the bankruptcy of the agent, he is no longer obliged to comply
with the contractual responsibilities and, if the derivative has a
negative value for the agent, then he draws a benefit, to the detriment
of the counterparty. Finally, the FVA, funding value adjustment, is
the change in value in the derivative that comes from the costs or
benefits, which the agent obtains following the collateralization
of the contract.

Recently, these issues have attracted the attention of many academics
and nowadays the literature on credit value adjustment is large. The
most common approach to XVA valuation consists in computing the price
of the contract subject to risk through a PDE. One of the first authors
to suggest a PDE based approach is Piterbarg \cite{piterbarg2010funding},
who introduces a model to include funding costs on derivative valuations
when collateral has to be posted. Burgard and Kjaer \cite{burgard2010pde}
propose a more general model for the evaluation of bilateral counterparty
risk and funding costs still based on the description of the value
of European-type derivatives in terms of PDE. De Graaf et al. \cite{de2016efficient,de2014efficient}
propose the so-called finite-difference Monte Carlo (FDMC) method,
which exploits both finite-difference and Monte Carlo methods to compute
the CVA and to compute first and second-order sensitivities for counterparty
credit risk. Feng \cite{feng2017cva} adapts the FDMC method to deal
with the case of an underlying evolving according to the Bates model:
in this particular case, the PDE to be solved is replaced by a partial
integral differential equation (PIDE), which implies an additional
computational effort. Goudenège et al. \cite{goudenege2020computing}
improve the method proposed by De Graaf et al. and compute the CVA
in the Bates model by solving coupled PIDEs. 

Other authors have considered the Monte Carlo method. Ballotta et
al. \cite{ballotta2019integrated} use Monte Carlo and Fourier transform
based methods to study a structural model when the underlying follows
a Lévy process. Brigo and Vrins \cite{brigo2018disentangling} use
Monte Carlo to evaluate CVA in a model that effectively manages wrong-way
risk. Antonelli et al. \cite{antonelli2022approximate} propose a
procedure based on a Taylor approximation for evaluating XVA and compare
it against Monte Carlo simulations.

Recently, Arregui et al. \cite{arregui2017pde,arregui2019monte} extend
the model of Burgard and Kjaer \cite{burgard2010pde} to the analysis
of American-type derivatives. This line of research is taken up by
Salvador and Oosterlee \cite{salvador2021total}, who develop the
stochastic model for the underlying by considering stochastic volatility.
Furthermore, Yuan et al. \cite{yuan2022total} present two different
numerical approaches to estimate the total value adjustments of the
Bermudan option, under the pure jump CGMY model. 

Numerical techniques for option pricing that rely solely on PDEs generally
suffer from the curse of dimensionality, that is the explosion of
computational cost in the presence of high dimensional problems, whereas
standard Monte Carlo methods are not effective in the case of American
options. The previously discussed methods for calculating XVA are
not exempt from this limitation. 

Newer techniques for evaluating derivatives make use of machine learning
methods. In this regard, some authors employ neural networks. For
example, Lapeyre and Lelong \cite{lapeyre2021neural} study the Longstaff
and Schwartz algorithm when the standard least-square regression is
replaced by a neural network approximation. Becker et al. \cite{becker2019deep,becker2020pricing,becker2021solving}
develop deep learning methods for pricing and hedging American-style
and, more generally, for solving optimal control problems. Other authors
exploit Gaussian Process Regression (GPR), a Machine Learning technique
that allows for estimations from scattered data in large dimensional
spaces. In this regard, we mention the work of Ludkovski \cite{ludkovski2018},
who evaluates Bermudan options by fitting the continuation values
through GPR. More recently, Goudenège et al. \cite{goudenege2020machine}
propose three GPR-based algorithm, termed GPR-MC, GPR-Tree and GPR-EI,
for pricing American options on a basket of assets following multi-dimensional
Black-Scholes dynamics.

The literature on the computation of the XVA for high dimensional
derivatives is rather sparse. As far as the computation of the credit
adjustments are concerned, She and Grecu \cite{she2018neural} compute
CVA and DVA by employing neural network as a universal approximator.
Cr{\'e}pey and Dixon \cite{crepey2019gaussian} exploit GPR to speed
up the computation of the CVA for derivatives portfolios. Gnoatto
et al. \cite{gnoatto2020deep} exploit artificial neural network to
compute the XVA for large portfolios of derivatives. Despite the importance
of this topic, to our knowledge, no one has ever studied the calculation
of XVA for American basket options, which are probably the most popular
option involving several assets. 

In this paper, we aim to fill this gap, by proposing an approach based
on a suitable probabilistic formulation of the XVA, derived from the
model of Burgard and Kjaer \cite{burgard2010pde}, which exploits
the GPR-MC and the GPR-EI algorithms for option pricing to overcome
the curse of dimensionality. We point out that we have chosen to consider
Burgard and Kjaer's model as it is particularly suitable as, unlike
other models, the American option exercise strategy is shaped to take
into account the probability of default of any of the agents. Moreover,
depending on the choice of the mark-to market value, two possible
kinds of models are considered: a linear and a non-linear. Furthermore,
the computation accuracy is increased by exploiting suitable control
variate for both the riskless and the risky price. Numerous numerical
tests demonstrate the reliability and accuracy of the proposed procedures
when different derivatives are considered.

The remainder of the paper is organized as follows. In Section 2 we
introduce the model for XVA on American basket options. In Section
3 we describe the proposed procedures. In Section 4 we discuss numerical
results. Finally, in Section 5, we conclude.

\section{Total value adjustment for American basket options}

In this Section, we describe a PDE-based model for the total value
adjustment when American options are concerned and we discuss a probabilistic
interpretation that we are going to exploit for our approaches. We
stress out that the model we develop here is inspired by the framework
previously introduced by Burgard and Kjaer \cite{burgard2010pde}
and developed by Arregui et al. \cite{arregui2017pde,arregui2019monte},
which is very interesting among the others because it allows the exercise
strategy of the American option to be influenced by the possibility
of default of each agent. This phenomenon, which is certainly plausible
in reality, is not present in other models. For example, the model
by De Graaf et al. \cite{de2014efficient,de2016efficient}, which
is usually employed be other authors, considers the strategy for the
risky option to be the same as the strategy for options without default
risk but, in our opinion, this does not seem to be the right choice.
This aspect is pointed out in the following Remark.
\begin{rem}
\emph{Consider an American call option, which is at the money at the
time of issue. Now, suppose an agent buys such an option from a counterparty
that provides a null recovery rate and that is going to default before
the maturity of the option almost sure. It is well known that if the
underlying does not pay dividends, and the counterparty does not default,
it is never optimal to exercise an American call option before maturity.
So, if the agent employs the standard strategy, he will achieve a
payoff equal to zero almost sure, as the default of the counterparty
will occur before maturity and the option will lose all its value
(the recovery rate is zero). On the other hand, if he exercises the
option immediately, he will obtain a positive payoff (the option is
in-the-money), so this strategy is better than the classical one.
This simple example shows that the optimal strategy for exercising
an American option must take the default risk into account.}
\end{rem}

Let 
\[
\mathbf{S}=(\mathbf{S}_{t})_{t\in[0,T]}=\left(S_{t}^{1},\dots,S_{t}^{d}\right)_{t\in[0,T]}
\]
 denote a $d$-dimensional stochastic process following the multi-dimensional
Black-Scholes model. Under the risk neutral probability $\mathbb{Q}$,
the dynamics of each underlying is given by
\begin{equation}
dS_{t}^{i}=\left(r-\eta_{i}\right)\,S_{t}^{i}\,dt+\sigma_{i}\,S_{t}^{i}\,dW_{t}^{i},\quad\ i=1,\ldots,d,\label{sde}
\end{equation}
with $\mathbf{S}_{0}=\left(s_{0}^{1},\dots,s_{0}^{d}\right)\in\mathbb{R}_{+}^{d}$
the spot price, $r$ the (constant) interest rate, $\eta=(\eta_{1},\dots,\eta_{d})$
 the vector of dividend rates, $\mathbf{\sigma}=(\sigma_{1},\dots,\sigma_{d})$
 the vector of volatilities, $\mathbf{W}$ a $d$-dimensional correlated
Brownian motion and $\rho_{ij}$ the instantaneous correlation coefficient
between $W_{t}^{i}$ and $W_{t}^{j}.$ 

Let us consider an American option issued at time $0$ with maturity
$T$ and let $H\,:\,\R_{+}^{d}\to\R$ denote the payoff function.
Let us term B the issuer and C the buyer of the option. For the moment,
we suppose that none of the two agents can default. We approximate
the value of the risk-less American option by a Bermudan option which
can be exercise at the times $t_{n}=n\cdot\Delta t$ for $n=0,\dots,N$
with $\Delta t=\nicefrac{T}{N}$ and $N\in\mathbb{N}.$ By employing
standard arguments, one can prove that
\[
V\left(t_{n},\mathbf{S}_{t}\right)=\max\left(C\left(t,\mathbf{\mathbf{S}_{t}}\right),H\left(\mathbf{S}_{t}\right)\right),
\]
where $C\left(t,\mathbf{\mathbf{S}_{t}}\right)$ stands for the continuation
value. In particular, $C\left(t,\mathbf{\mathbf{S}_{t}}\right)$ restricted
to the time interval $\left]t_{n},t_{n+1}\right[$ is equal to $C^{n}\left(t,\mathbf{\mathbf{S}_{t}}\right)$,
the solution of the following PDE, defined in $\left]t_{n},t_{n+1}\right[$for
$n=0,\dots,N-1$ :
\[
\frac{\partial C^{n}}{\partial t}+\mathcal{A}\left(C^{n}\right)-rC^{n}=0,
\]
with $C^{n}=C^{n}\left(t_{n+1},\mathbf{x}\right)$ for $\mathbf{x}=\left(\mathbf{x}_{1},\dots,\mathbf{x}_{d}\right)$
and
\[
\mathcal{A}\left(C^{n}\right)=\sum_{i=1}^{d}\left(r-\eta_{i}\right)\mathbf{x}_{i}\frac{\partial C^{n}}{\partial\mathbf{x}_{i}}+\sum_{i=1}^{d}\frac{\sigma_{i}^{2}\mathbf{x}_{i}^{2}}{2}\frac{\partial^{2}C^{n}}{\partial\mathbf{x}_{i}^{2}}+\sum_{i=1}^{d-1}\sum_{j=i+1}^{d}\rho_{i,j}\sigma_{i}\sigma_{j}\mathbf{x}_{i}\mathbf{x}_{j}\frac{\partial^{2}C^{n}}{\partial\mathbf{x}_{i}\partial\mathbf{x}_{j}}.
\]
The terminal condition is 
\[
C^{n}\left(t_{n+1},\mathbf{x}\right)=\begin{cases}
H\left(\mathbf{x}\right) & \text{if }n+1=N,\\
V\left(t_{n+1},\mathbf{x}\right) & \text{otherwise}.
\end{cases}
\]

Now, let us suppose that both agents B and C can default. We take
the point of B, and we denote the risky option price by $\hat{V}\left(t,\mathbf{S}_{t},J_{t}^{B},J_{t}^{C}\right)$,
with $J^{B}$ and $J^{C}$ two independent jump processes that change
value, from 0 to 1, at the time the corresponding agent defaults.

Let $M_{t}=M\left(t,\mathbf{S}_{t}\right)$ represent the close-out
mark-to-market value, that is, the monetary value of the contract
used as the basis for settlement. Let us define $M^{+}=\max\left(M,0\right)$
and $M^{-}=\min\left(M,0\right)$. Following Burgard and Kjaer \cite{burgard2010pde},
in case of default of one counterparty, the risky values are defined
as follows: 
\begin{itemize}
\item if the issuer B defaults first, 
\[
\hat{V}\left(t,\mathbf{S}_{t},1,0\right)=M_{t}^{+}+R_{B}M_{t}^{-},
\]
with $R_{B}\in[0,1]$ the recovery rate of C respect to the default
of B;
\item if the buyer C defaults first, 
\[
\hat{V}\left(t,\mathbf{S}_{t},0,1\right)=R_{C}M_{t}^{+}+M_{t}^{-},
\]
with $R_{C}\in[0,1]$ the recovery rate of B respect to the default
of C.
\end{itemize}
Let $\lambda_{B}$ and $\lambda_{C}$ be the constant default intensities
of B and C, respectively, and $s_{F}$ the funding cost of B. According
to Burgard and Kjaer \cite{burgard2010pde}, if the derivative can
be used as a collateral, then $s_{F}=0$, and if it cannot, then $s_{F}=\left(1-R_{B}\right)\lambda_{B}$.
Following Arregui et al. \cite{arregui2017pde,arregui2019monte},
the value $\hat{V}\left(t,\mathbf{S}_{t},0,0\right)$ of the Bermudan
risky option, satisfies 
\[
\hat{V}\left(t,\mathbf{S}_{t},0,0\right)=\max\left(\hat{C}\left(t,\mathbf{\mathbf{S}_{t}}\right),H\left(\mathbf{S}_{t}\right)\right),
\]
with $\hat{C}$ the continuation value of the risky option. Similarly
to what happens for the risk-free option, $\hat{C}\left(t,\mathbf{\mathbf{S}_{t}}\right)$
restricted to the time interval $\left]t_{n},t_{n+1}\right[$ is equal
to $\hat{C}^{n}\left(t,\mathbf{\mathbf{S}_{t}}\right)$, the solution
of the following PDE, defined in $\left]t_{n},t_{n+1}\right[$ for
$n=0,\dots,N-1$ :
\begin{equation}
\frac{\partial\hat{C}^{n}}{\partial t}+\mathcal{A}\left(\hat{C}^{n}\right)-r\hat{C}^{n}=\left(\lambda_{B}+\lambda_{C}\right)\hat{C}^{n}+s_{F}M^{+}-\lambda_{B}\left(R_{B}M^{-}+M^{+}\right)-\lambda_{C}\left(R_{C}M^{+}+M^{-}\right),\label{eq:Vnhat}
\end{equation}
 with the terminal condition
\[
\hat{C}^{n}\left(t_{n+1},\mathbf{x}\right)=\begin{cases}
H\left(\mathbf{x}\right) & \text{if }n+1=N,\\
\hat{V}\left(t_{n+1},\mathbf{x},0,0\right) & \text{otherwise}.
\end{cases}
\]

We proceed backward in time. Suppose $\hat{V}^{n}\left(t_{n+1},\mathbf{x}\right)$
is known and we aim to compute $\hat{V}^{n}\left(t_{n},\mathbf{x}\right)$.
By the Feynman-Kac formula applied to equation (\ref{eq:Vnhat}),
(see e.g. Platen and Heath \cite{platen2006benchmark}), we have

\begin{equation}
\hat{C}^{n}\left(t_{n},\mathbf{x}\right)=\mathbb{E}^{\mathbb{Q}}\left[\int_{t_{n}}^{t_{n+1}}e^{-r_{0}\left(u-t_{n}\right)}g\left(u,\mathbf{S}_{u}\right)du+e^{-r_{0}\Delta t}\hat{C}^{n}\left(t_{n+1},\mathbf{S}_{t_{n+1}}\right)\mid S_{t_{n}}=\mathbf{x}\right],\label{eq:backward}
\end{equation}
with 
\[
r_{0}=r+\lambda_{B}+\lambda_{C},
\]
\begin{align*}
g\left(u,\mathbf{S}_{u}\right) & =-\left[s_{F}M_{u}^{+}-\lambda_{B}\left(R_{B}M_{u}^{-}+M_{u}^{+}\right)-\lambda_{C}\left(R_{C}M_{u}^{+}+M_{u}^{-}\right)\right]\\
 & =M_{u}^{+}\left(\lambda_{B}+\lambda_{C}R_{C}-s_{F}\right)+M_{u}^{-}\left(\lambda_{C}+\lambda_{B}R_{B}\right)\\
 & =M_{u}^{+}c_{p}+M_{u}^{-}c_{m},
\end{align*}
 $c_{p}=\lambda_{B}+\lambda_{C}R_{C}-s_{F}$ and $c_{m}=\lambda_{C}+\lambda_{B}R_{B}$.
In particular, as $\lambda_{B},\lambda_{C},R_{C}$ and $R_{B}$ are
positive quantities and $s_{F}$ is equal to $0$ or $\left(1-R_{B}\right)\lambda_{B}$,
thus $c_{p}$ and $c_{m}$ are positive values.

We approximate the integral in (\ref{eq:backward}) by a two points
trapezoidal quadrature rule:

\begin{align*}
\hat{C}^{n}\left(t_{n,},\mathbf{x}\right) & \approx\mathbb{E}^{\mathbb{Q}}\left[\frac{e^{-r_{0}\Delta t}g\left(t_{n+1,},\mathbf{S}_{t_{n+1}}\right)+g\left(t_{n,},\mathbf{S}_{t_{n}}\right)}{2}\Delta t+e^{-r_{0}\Delta t}\hat{C}^{n}\left(t_{n+1},\mathbf{S}_{t_{n+1}}\right)\mid S_{t_{n}}=\mathbf{x}\right]\\
 & =e^{-r_{0}\Delta t}\mathbb{E}^{\mathbb{Q}}\left[\frac{\Delta t}{2}g\left(t_{n+1,},\mathbf{S}_{t_{n+1}}\right)+\hat{C}^{n}\left(t_{n+1},\mathbf{S}_{t_{n+1}}\right)\mid S_{t_{n}}=\mathbf{x}\right]+\frac{\Delta t}{2}g\left(t_{n,},\mathbf{S}_{t_{n}}\right),
\end{align*}
and thus

\begin{equation}
\hat{V}\left(t_{n,},\mathbf{x},0,0\right)\approx\max\left\{ e^{-r_{0}\Delta t}\mathbb{E}^{\mathbb{Q}}\left[\frac{\Delta t}{2}g\left(t_{n+1,},\mathbf{S}_{t_{n+1}}\right)+\hat{V}\left(t_{n+1},\mathbf{S}_{t_{n+1}},0,0\right)\mid S_{t_{n}}=S\right]+\frac{\Delta t}{2}g\left(t_{n,},\mathbf{x}\right),H\left(\mathbf{x}\right)\right\} \label{eq:expectation}
\end{equation}
Now, we distinguish two cases: $M_{u}=V\left(u,S_{u}\right)$, that
is the value of the risk-free derivative, and $M_{u}=\hat{V}\left(u,S_{u},0,0\right),$
that is the value of the defaultable derivative. 

\subsection{Case $M=V$}

We suppose that the values of $V$ have already been computed in a
suitable domain. If $M_{u}=V\left(u,\mathbf{S}_{u}\right)$, we can
compute $\hat{V}\left(t_{n,},\mathbf{S}_{t_{n,}},0,0\right)$ explicitly,
by replacing $M$ with the pre-computed values of $V$ and by approximating
the expectation in (\ref{eq:expectation}) by a suitable numeric technique.

\subsection{Case $M=\hat{V}$ }

If $M_{u}=\hat{V}\left(u_{,},S_{u}\right)$, then

\begin{equation}
\hat{V}\left(t_{n,},\mathbf{x},0,0\right)\approx\max\left\{ E\left(\mathbf{x}\right)+\frac{\Delta t}{2}\left(\hat{V}\left(t_{n,},\mathbf{x},0,0\right)^{+}c_{p}+\hat{V}\left(t_{n,},\mathbf{x},0,0\right)^{-}c_{m}\right),H\left(\mathbf{x}\right)\right\} ,\label{eq:implicitVhat}
\end{equation}
with 
\begin{equation}
E\left(\mathbf{x}\right)=e^{-r_{0}\Delta t}\mathbb{E}^{\mathbb{Q}}\left[\frac{\Delta t}{2}\left(\hat{V}\left(t_{n+1},\mathbf{S}_{t_{n+1}},0,0\right)^{+}c_{p}+\hat{V}\left(t_{n+1},\mathbf{S}_{t_{n+1}},0,0\right)^{-}c_{m}\right)+\hat{V}\left(t_{n+1},\mathbf{S}_{t_{n+1}},0,0\right)\mid S_{t_{n}}=\mathbf{x}\right].\label{eq:EXP2}
\end{equation}
We define $\tilde{V}\left(t_{n,},\mathbf{x}\right)$ as the solution
of the implicit equation problem 
\begin{equation}
\tilde{V}\left(t_{n,},\mathbf{x}\right)=\max\left\{ E\left(\mathbf{x}\right)+\frac{\Delta t}{2}\left(\tilde{V}\left(t_{n,},\mathbf{x}\right)^{+}c_{p}+\tilde{V}\left(t_{n,},\mathbf{x}\right)^{-}c_{m}\right),H\left(\mathbf{x}\right)\right\} ,\label{eq:implicitVtilde}
\end{equation}
and we employ it as an approximation of $\hat{V}\left(t_{n,},\mathbf{x},0,0\right)$.
Equation (\ref{eq:implicitVtilde}) is implicit -- $\hat{V}$ appears
both on left and the right side of the equation -- and non linear.
The following proposition discuss how to solve it.

\begin{Prop}\label{Prop1} Let $\tilde{V}\left(t_{n,},\mathbf{x}\right)$
be the unique solution of the implicit equation (\ref{eq:implicitVtilde}).
Then, if $H\left(\mathbf{x}\right)\leq0$:
\begin{itemize}
\item if $E\left(\mathbf{x}\right)\leq H\left(\mathbf{x}\right)\left(1-\frac{\Delta t}{2}c_{m}\right)\leq0$
then $\tilde{V}\left(t_{n,},\mathbf{x}\right)=H\left(\mathbf{x}\right)$;
\item if $H\left(\mathbf{x}\right)\left(1-\frac{\Delta t}{2}c_{m}\right)<E\left(\mathbf{x}\right)\leq0$
then $\tilde{V}\left(t_{n,},\mathbf{x}\right)=\frac{E\left(\mathbf{x}\right)}{1-\frac{\Delta t}{2}c_{m}}$;
\item if $E\left(\mathbf{x}\right)>0$ then $\tilde{V}\left(t_{n,},\mathbf{x}\right)=\frac{E\left(\mathbf{x}\right)}{1-\frac{\Delta t}{2}c_{p}}$.
\end{itemize}
If $H\left(\mathbf{x}\right)>0$: 
\begin{itemize}
\item if $E\left(\mathbf{x}\right)\leq H\left(\mathbf{x}\right)\left(1-\frac{\Delta t}{2}c_{p}\right)$
then $\tilde{V}\left(t_{n,},\mathbf{x}\right)=H\left(\mathbf{x}\right)$;
\item if $E\left(\mathbf{x}\right)>H\left(\mathbf{x}\right)\left(1-\frac{\Delta t}{2}c_{p}\right)$
then $\tilde{V}\left(t_{n,},\mathbf{x}\right)=\frac{E\left(\mathbf{x}\right)}{1-\frac{\Delta t}{2}c_{p}}$.
\end{itemize}
\end{Prop}

The proof of Proposition \ref{Prop1} is discussed in the Appendix
\ref{App1}.

\section{Gaussian Process Regression for computing XVA}

According to the previous Section, the calculation of XVA requires
the computation of an expected value, both in the case $M=V$ and
in the case $M=\hat{V}$, see (\ref{eq:expectation}). This calculation
involves a stochastic underlying which is a multidimensional process,
potentially high dimensional. We propose to use two techniques, already
successfully applied by Goudenège et al. \cite{goudenege2020machine}
for multidimensional option pricing problems: GPR-MC and GPR-EI. 

Below, we recall the main aspects of these two methods, and we refer
the interested reader to \cite{goudenege2020machine} for more information.

\subsection{GPR-MC}

The GPR Monte Carlo approach employs Monte Carlo simulations to compute
the continuation value of a Bermudan option and GPR to learn the option
value at each time step.

The algorithm starts by simulating a set of trajectories of the underlyings.
Let $X^{n}$ represent the set of $P$ points whose coordinates represent
certain possible values for the underlyings at time $t_{n}$, for
$n=0,\dots,N,$ that is 
\begin{equation}
X^{n}=\left\{ \mathbf{x}^{n,p}=\left(x_{1}^{n,p},\dots,x_{d}^{n,p}\right),p=1,\dots,P\right\} \subset\mathbb{R}^{d}.
\end{equation}
The points of the sets $X^{n}$ are computed by employing the Halton's
low-discrepancy sequence in $\mathbb{R}^{d}$ and standard algorithms
for simulating the undelying values in the multidimensional Black-Scholes
model. 

Now, suppose we want to compute the continuation value of an Bermudan
option but only for $\mathbf{S}_{t_{n}}=\mathbf{x}^{n,p}\in X^{n}$.
This goal can be achieved by means of a one step Monte Carlo simulation.
In particular, for each $\mathbf{x}^{n,p}\in X^{n}$, we simulate
a set of $M$ points 
\[
\tilde{X}_{p}^{n}=\left\{ \mathbf{\tilde{x}}^{n,p,m}=\left(\tilde{x}_{1}^{n,p,m},\dots,\tilde{x}_{d}^{n,p,m}\right),m=1,\dots,M\right\} \subset\mathbb{R}^{d},
\]
which are possible values for $\mathbf{S}_{t_{n+1}}$ according to
the law of $\mathbf{S}_{t_{n+1}}\left|\ensuremath{\mathbf{S}_{t_{n}}=\mathbf{x}^{n,p}}\right.$.
In particular, for $i=1,\dots,d$, $n=1,\dots,N$, $p=1,\dots,P$,
$m=1,\dots,M$, we define 
\begin{equation}
\tilde{x}_{i}^{n,p,m}=x_{i}^{n,p}e^{\left(r-\eta_{i}-\frac{1}{2}{\sigma}_{i}^{2}\right)\Delta t+\sqrt{\Delta t}\sigma_{i}\Sigma_{i}\mathbf{G}^{n,p,m}},\label{xtilde}
\end{equation}
where $\mathbf{G}^{n,p,m}\sim\mathcal{N}\left(0,I_{d}\right)$ is
a standard Gaussian random vector and $\Sigma_{i}$ is the $i$-th
row of the matrix $\Sigma$, which is defined as a square root of
the correlation matrix $\Gamma$ of the multidimensional Brownian
increments. Thus, the risk-less option value can be approximated for
each $\mathbf{x}^{n,p}\in X^{n}$ by the following scheme:
\begin{equation}
\begin{cases}
V_{n}^{MC}\left(\mathbf{x}^{n,p}\right)=\max\left(\frac{e^{-r\Delta t}}{M}\sum_{m=1}^{M}V_{n+1}^{MC}\left(\mathbf{\tilde{x}}^{n,p,m}\right),H\left(\mathbf{x}^{n,p}\right)\right) & \text{if }n<N,\\
V^{MC}\left(t_{n},\mathbf{x}^{n,p}\right)=H\left(\mathbf{x}^{n,p}\right) & \text{if }n=N.
\end{cases}\label{eq:update2}
\end{equation}
Furthermore, the risky value $\hat{V}_{n}^{MC}\left(\mathbf{x}^{n,p}\right)$
for $M=V$, $t=t_{n}$ and $\mathbf{S}_{t_{n}}=\mathbf{x}^{n,p}$
is approximated by the following scheme:
\begin{equation}
\begin{cases}
\hat{V}_{n}^{MC}\left(\mathbf{x}^{n,p}\right)=\max\left\{ \hat{C}_{n}^{MC}\left(\mathbf{x}^{n,p}\right),H\left(\mathbf{x}^{n,p}\right)\right\}  & \text{if }n<N,\\
\hat{V}_{n}^{MC}\left(\mathbf{x}^{n,p}\right)=H\left(\mathbf{x}^{n,p}\right) & \text{if }n=N,
\end{cases}\label{update3}
\end{equation}
with 
\begin{multline*}
\hat{C}_{n}^{MC}\left(\mathbf{x}^{n,p}\right)=\frac{e^{-r_{0}\Delta t}}{M}\sum_{m=1}^{M}\left[\frac{\Delta t}{2}\left(V_{n+1}^{MC}\left(\mathbf{\tilde{x}}^{n,p,m}\right)^{+}c_{p}+V_{n+1}^{MC}\left(\mathbf{\tilde{x}}^{n,p,m}\right)^{-}c_{m}\right)+\hat{V}_{n+1}^{MC}\left(\mathbf{\tilde{x}}^{n,p,m}\right)\right]+\\
+\frac{\Delta t}{2}\left(V_{n}^{MC}\left(\mathbf{x}^{n,p}\right)^{+}c_{p}+V_{n}^{MC}\left(\mathbf{x}^{n,p}\right)^{-}c_{m}\right).
\end{multline*}
Finally, the risky value $\hat{V}_{n}^{MC}\left(\mathbf{x}^{n,p}\right)$,
for $M=\hat{V}$, $t=t_{n}$ and $\mathbf{S}_{t_{n}}=\mathbf{x}^{n,p}$,
is computed according to Proposition \ref{Prop1}, with $E\left(\mathbf{x}^{n,p}\right)$
approximated by 
\begin{equation}
E^{MC}\left(\mathbf{x}^{n,p}\right)=\frac{e^{-r_{0}\Delta t}}{M}\sum_{m=1}^{M}\left[\frac{\Delta t}{2}\left(\hat{V}_{n+1}^{MC}\left(\mathbf{\tilde{x}}^{n,p,m}\right)^{+}c_{p}+\hat{V}_{n+1}^{MC}\left(\mathbf{\tilde{x}}^{n,p,m}\right)^{-}c_{m}\right)+\hat{V}_{n+1}^{MC}\left(\mathbf{\tilde{x}}^{n,p,m}\right)\right].\label{eq:update4}
\end{equation}

If we proceed backward, the functions $V_{N}^{MC}$ and $\hat{V}_{N}^{MC}$
are known since they are equal to the payoff of the option $H$, so
one can compute both $V_{N-1}^{MC}$ and $\hat{V}_{N-1}^{MC}$ at
$\tilde{X}_{p}^{n}$ by exploiting equations (\ref{eq:update2}),
(\ref{update3}) or (\ref{eq:update4}). Similarly, such a computation
at a time step $t_{n}$ with $n<N$ requires the knowledge of the
value functions $V_{n+1}^{MC}$ and $\hat{V}_{n+1}^{MC}$ at the next
time step $t_{n+1}$ at all the points of the set $\bigcup_{p=1,\dots,P}\tilde{X}_{p}^{n+1}$,
but, following the procedure just described, those functions are known
only at the points of the set $X^{n+1}$: a multidimensional extrapolation
tool is required to extending the value functions from $X^{n}$ to
a suitable neighbourhood of such a set. For this purpose, we exploit
Gaussian Process Regression, a class of non-parametric kernel-based
probabilistic models that represents the input data as the random
observations of a Gaussian stochastic process and it employs a Bayesian
approach to perform estimation of the process at new input data. This
Machine Learning techniques is well suited to our problem, as it is
capable of handling randomly scattered input data and, generally,
only a few input observations are needed to obtain accurate predictions.
For a brief introduction to GPR, we refer the interested reader to
De Spiegeleer et al. \cite{de2018machine} or to Goudenège et al.
\cite{goudenege2020machine}, while for a more in-depth discussion,
we suggest Rasmussen and Williams \cite{rasmussen2006gaussian}.

Let $V_{n}^{GPR-MC}$ and $\hat{V}_{n}^{GPR-MC}$ be the GPR approximations
of the functions $V_{n}^{MC}$ and $\hat{V}_{n}^{MC}$, obtained from
the observations $\left\{ \left(\mathbf{x}^{n,p},V_{n}^{MC}\left(\mathbf{x}^{n,p}\right)\right),p=1,\dots,P\right\} $
and $\left\{ \left(\mathbf{x}^{n,p},\hat{V}_{n}^{MC}\left(\mathbf{x}^{n,p}\right)\right),p=1,\dots,P\right\} $
respectively. The GPR-MC algorithm requires the replacement of $V_{n+1}^{MC}$
and $\hat{V}_{n+1}^{MC}$ in the right side on (\ref{eq:update2}),
(\ref{update3}) or (\ref{eq:update4}) with $V_{n+1}^{GPR-MC}$ and
$\hat{V}_{n+1}^{GPR-MC}$ respectively.

\subsection{GPR-EI}

The GPR-Exact Integration method is similar to the GPR-MC method but
the continuation value is estimated through an exact computation of
the expectation, based on the Gaussian distribution. By contrast with
the GPR-MC method, the predictors employed in the GPR step are related
to the logarithms of the underlyings. Secondly, the continuation value
at these points is computed through a closed formula which comes from
an exact integration. 

Here, for the sake of brevity, we limit ourselves to pointing out
the main elements of this algorithm, and we refer the interested readers
to \cite{goudenege2020machine}. The computation of the continuation
value, for both risky or riskless options, is a particular case of
the computation of an expectation as
\[
\mathbb{E}^{\mathbb{Q}}\left[\Psi\left(\mathbf{S}_{t+\tau}\right)|\mathbf{S}_{t}=\mathbf{x}\right],
\]
with $\Psi$ a certain function, $t,t+\tau\in\left[0,T\right]$ and
$\tau>0$.

Let us define the input set
\[
Z=\left\{ \mathbf{z}^{p},p=1,\dots,P\right\} 
\]
 consisting of $P$ points in $\mathbb{R}^{d}$ quasi-randomly distributed
according to the law of the vector $\left(\sigma_{1}W_{\tau}^{1},\dots,\sigma_{d}W_{\tau}^{d}\right)^{\top}$.
In particular, we define 
\begin{equation}
\mathbf{z}_{i}^{p}=\sqrt{\tau}\sigma_{i}\Sigma_{i}\mathbf{h}^{p},
\end{equation}
 where $\Sigma_{i}$ is i-th row of the matrix $\Sigma$ and $\mathbf{h}^{p}$
is the q-th point of the Halton's low-discrepancy sequence in $\mathbb{R}^{d}$.
Let $u:Z\rightarrow\mathbb{R}$ be the function defined by 
\begin{equation}
u\left(\mathbf{z}\right):=\Psi\left(\mathbf{x}\exp\left(\left(r-\boldsymbol{\eta}-\frac{1}{2}\boldsymbol{\sigma}^{2}\right)\tau+\mathbf{z}\right)\right).\label{eq:u_def}
\end{equation}
The first step is to approximate the function $u$ by training the
GPR method with a Squared Exponential kernel on the set $Z$, so that
the GPR approximation of the function $u$ is given by
\begin{equation}
u^{GPR}\left(\mathbf{z}\right)=\sum_{p=1}^{P}k_{SE}\left(\mathbf{z}^{q},\mathbf{z}\right)\mathbf{\omega}_{p},
\end{equation}
where $\omega_{1},\dots,\omega_{P}$ are weights. The continuation
value can be computed by integrating the function $u^{GPR}$ against
a $d$-dimensional probability density. The use of the Squared Exponential
kernel allows one to easily perform such a calculation by means of
a closed formula, that is:

\begin{equation}
\mathbb{E}^{\mathbb{Q}}\left[\Psi\left(\mathbf{S}_{t+\tau}\right)|\mathbf{S}_{t}=\mathbf{x}\right]\approx\sum_{p=1}^{P}\omega_{q}\sigma_{f}^{2}\sigma_{l}^{d}\frac{e^{-\frac{1}{2}\left(\mathbf{z}^{p}\right)^{\top}\left(\tau\cdot\Pi+\sigma_{l}^{2}I_{d}\right)^{-1}\left(\mathbf{z}^{p}\right)}}{\sqrt{\det\left(\tau\cdot\Pi+\sigma_{l}^{2}I_{d}\right)}},\label{vEU1}
\end{equation}
 where $\sigma_{f}$, $\sigma_{l}$, and $\omega_{1},\dots,\omega_{Q}$
are certain constants determined by the GPR approximation of the function
$\mathbf{z}\mapsto u\left(\mathbf{z}\right)$ considering $Z$ as
the predictor set, and $\Pi=\left(\Pi_{i,j}\right)$ is the $d\times d$
covariance matrix of the vector $\left(\sigma_{1}W_{T}^{1},\dots,\sigma_{d}W_{T}^{d}\right)^{\top}$,
that is $\Pi_{i,j}=\rho_{i,j}\sigma_{i}\sigma_{j}$.

\subsection{Control Variates}

As suggested by Goudenège et al. \cite{goudenege2021variance}, control
variates technique is a usefull tool to improve the accuracy of pricing
methods based on GRP. Specifically, we use the European risk-less
price $V^{EU}$ as the control variate for the American risk-less
price, and the American risk-less price for the American risky price.
In particular, we compute the European risk-less price by Monte Carlo
simulations with antithetic variates. We explain the use of control
variates technique for the computation of the risk-less American option
price $V$, and we leave the appropriate adjustments for the risky
price $\hat{V}$ to the reader. 

Let $V^{EU}$ represent the risk-less price of the European option.
For a fixed time $t$ and an underlying stocks value $\mathbf{x}$,
the American-European price gap is defined as the difference between
the American and the European price, that is: 
\begin{equation}
v\left(t,\mathbf{x}\right)=V\left(t,\mathbf{x},0,0\right)-V^{EU}\left(t,\mathbf{x},0,0\right).\label{eq:AM_EU_gap}
\end{equation}
 The price gap is equal to zero at maturity and, at a general time
$t$, it can be computed as
\begin{equation}
v\left(t,\mathbf{x}\right)=\sup_{\tau\in\mathcal{T}_{t,T}}\mathbb{E}^{\mathbb{Q}}\left[e^{-r\left(\tau-t\right)}K\left(\tau,\mathbf{S}_{\tau}\right)|\mathbf{S}_{t}=\mathbf{x}\right],
\end{equation}
 where $\mathcal{T}_{t,T}$ stands for the set of all stopping times
taking values in $\left[t,T\right]$ and $K$ is the exercise value
gap, defined by 
\begin{equation}
K\left(t,\mathbf{x}\right)=H\left(\mathbf{x}\right)-V^{EU}\left(t,\mathbf{x},0,0\right).\label{Psihat}
\end{equation}
 Therefore, the function $v\left(t,\mathbf{x}\right)$ can be estimated
by exploiting a dynamic programming principle based on Bermudan approximation.
In particular, one can use GPR-MC and GPR-EI, by replacing $H$ with
$K$. Finally, after computing the initial price gap $v\left(0,\mathbf{S}_{0}\right)$,
by inverting relation (\ref{eq:AM_EU_gap}), one can obtain the American
price as
\begin{equation}
V\left(0,\mathbf{S}_{0},0,0\right)=v\left(0,\mathbf{S}_{0}\right)+V^{EU}\left(0,\mathbf{S}_{0},0,0\right).
\end{equation}

\begin{rem}
\emph{The computation of the European prices for the control variates
technique and the expectation (\ref{eq:expectation}) are the most
time demanding steps. However, these steps can easily be parallelised,
thus reducing the total computational time.}
\end{rem}

\section{Numerical experiments}

In this Section we propose the results of some numerical experiments.
The algorithms have been implemented in MATLAB and computations have
been preformed on a server which employs a $2.40$ GHz Intel$^{\circledR}$
Xeon$^{\circledR}$ processor (Gold 6148, Skylake) and 64 GB of RAM.
In the remainder of this Section, we discuss 3 American derivatives:
a Geometric Put, a Call on the maximum and a Swaption with floor.
Table \ref{tab:BS} lists all the parameters of the stochastic model,
with the exception of the dimension $d$, which takes on different
values from $d=2$ up to $d=80$. Based on the results discussed in
this Section, one can observe that the two proposed methods are very
accurate in the various cases considered. The quality of the results
degrades slightly as the size of the problem increases, but the quality
of the results is still acceptable, successfully limiting the effects
of the curse of dimensionality. Overall, the results proposed by the
two methods are always in agreement and very close to the benchmark
(when available). 

Finally, we stress that obtaining accurate values (in terms of relative
error) for the XVA is not an easy task. The XVA is in fact obtained
as the difference between two prices that are usually very close to
each other. A small estimation error on prices can have a significant
weight in relative terms on their difference.

\begin{table}
\begin{centering}
\begin{tabular}{llccllc}
\toprule 
Symbol & Meaning & Value &  & Symbol & Meaning & Value\tabularnewline
\midrule
$S_{0}^{i}$ & initial spot value & $100$ &  & $T$ & maturity & $1.0$\tabularnewline
$r$ & risk free i.r. & $0.03$ &  & $\lambda_{B}=\lambda_{C}$ & default intensities & $0.04$\tabularnewline
$\eta_{i}$ & dividend rate & $0.00$ &  & $R_{B}=R_{C}$ & recovery rates & $0.3$\tabularnewline
$\sigma_{i}$ & volatility & $0.25$ &  & $sF$ & funding cost & $0.028$\tabularnewline
$\rho_{i,j}$ & correlation & $0.2$ &  & $K$ & strike price & $100$\tabularnewline
\bottomrule
\end{tabular}
\par\end{centering}
\caption{\label{tab:BS}Parameters employed for the numerical experiments in
the multi-dimensional Black-Scholes model. In particular, $s_{F}=\left(1-R_{B}\right)\lambda_{B}$.}
\end{table}

\subsection*{Geometric Put }

We start by considering a Geometric Put option, whose payoff is 
\[
H(\mathbf{S}_{T})=\left(K-\left(\prod_{i=1}^{d}S_{T}^{i}\right)^{\frac{1}{d}}\right)_{+}.
\]

This is a very particularly interesting case since the value of this
$d$-dimensional option is equal to the value an appropriate one dimensional
American Put option in the Black-Scholes model, as pointed out in
\cite{goudenege2020machine,goudenege2021variance}. So, by using one-dimensional
standard techniques, such as the CRR tree or a finite difference algorithm,
one can obtain very accurate prices for both risk and risk-less American
option. In particular, we compute the American benchmark by using
both the CRR model with $4000$ time steps and a PDE approach with
$4000$ time steps and $4000$ space steps. The obtained values with
these two algorithms are equal to three decimal places, so that they
can be assumed reliable. The Bermudan benchmark is computed as the
American one, but the option has only $41$ possible exercise dates,
that is $t_{0}=0$, $t_{1}=\nicefrac{1}{40}$, $\dots$, $t_{40}=1$.
The GPR-MC method employs $40$ time steps, $2000$ points and $10^{4}$
Monte Carlo simulations, while the GPR-EI method employs $40$ time
steps and $2000$ points.

\begin{table}
\begin{centering}
\begin{tabular}{cccccccc}
\toprule 
 &  & \multicolumn{3}{c}{Option prices} &  & \multicolumn{2}{c}{XVA}\tabularnewline
\cmidrule{3-5} \cmidrule{4-5} \cmidrule{5-5} \cmidrule{7-8} \cmidrule{8-8} 
 &  & Risk-free & \multicolumn{2}{c}{With default risk} &  &  & \tabularnewline
$d$ &  &  & $M=V$ & $M=\hat{V}$ &  & $M=V$ & $M=\hat{V}$\tabularnewline
\midrule 
\multicolumn{8}{l}{American benchmark}\tabularnewline
$2$ &  & $6.901$ & $6.659$ & $6.657$ &  & $0.242$ & $0.244$\tabularnewline
$10$ &  & $4.866$ & $4.689$ & $4.688$ &  & $0.177$ & $0.178$\tabularnewline
$20$ &  & $4.530$ & $4.364$ & $4.363$ &  & $0.166$ & $0.167$\tabularnewline
$40$ &  & $4.350$ & $4.190$ & $4.189$ &  & $0.160$ & $0.161$\tabularnewline
$80$ &  & $4.257$ & $4.100$ & $4.099$ &  & $0.157$ & $0.158$\tabularnewline
\midrule 
\multicolumn{8}{l}{Bermudan benchmark}\tabularnewline
$2$ &  & $\underset{\left(-0.09\%\right)}{6.895}$ & $\underset{\left(-0.12\%\right)}{6.651}$ & $\underset{\left(-0.12\%\right)}{6.649}$ &  & $\underset{\left(0.83\%\right)}{0.244}$ & $\underset{\left(0.82\%\right)}{0.246}$\tabularnewline
$10$ &  & $\underset{\left(-0.06\%\right)}{4.863}$ & $\underset{\left(-0.09\%\right)}{4.685}$ & $\underset{\left(-0.11\%\right)}{4.683}$ &  & $\underset{\left(0.56\%\right)}{0.178}$ & $\underset{\left(1.12\%\right)}{0.180}$\tabularnewline
$20$ &  & $\underset{\left(-0.07\%\right)}{4.527}$ & $\underset{\left(-0.09\%\right)}{4.360}$ & $\underset{\left(-0.11\%\right)}{4.358}$ &  & $\underset{\left(0.60\%\right)}{0.167}$ & $\underset{\left(1.20\%\right)}{0.169}$\tabularnewline
$40$ &  & $\underset{\left(-0.07\%\right)}{4.347}$ & $\underset{\left(-0.10\%\right)}{4.186}$ & $\underset{\left(-0.10\%\right)}{4.185}$ &  & $\underset{\left(0.63\%\right)}{0.161}$ & $\underset{\left(1.24\%\right)}{0.163}$\tabularnewline
$80$ &  & $\underset{\left(-0.07\%\right)}{4.254}$ & $\underset{\left(\text{-0.10\%}\right)}{4.096}$ & $\underset{\left(-0.10\%\right)}{4.095}$ &  & $\underset{\left(0.64\%\right)}{0.158}$ & $\underset{\left(0.63\%\right)}{0.159}$\tabularnewline
\midrule 
\multicolumn{8}{l}{GPR-MC}\tabularnewline
$2$ &  & $\underset{\left(-0.10\%\right)}{6.894}$ & $\underset{\left(-0.14\%\right)}{6.650}$ & $\underset{\left(-0.14\%\right)}{6.648}$ &  & $\underset{\left(0.83\%\right)}{0.244}$ & $\underset{\left(0.82\%\right)}{0.246}$\tabularnewline
$10$ &  & $\underset{\left(-0.04\%\right)}{4.864}$ & $\underset{\left(-0.09\%\right)}{4.685}$ & $\underset{\left(-0.09\%\right)}{4.684}$ &  & $\underset{\left(1.13\%\right)}{0.179}$ & $\underset{\left(1.69\%\right)}{0.181}$\tabularnewline
$20$ &  & $\underset{\left(0.00\%\right)}{4.530}$ & $\underset{\left(-0.09\%\right)}{4.360}$ & $\underset{\left(-0.09\%\right)}{4.359}$ &  & $\underset{\left(1.81\%\right)}{0.169}$ & $\underset{\left(2.40\%\right)}{0.171}$\tabularnewline
$40$ &  & $\underset{\left(0.00\%\right)}{4.350}$ & $\underset{\left(-0.07\%\right)}{4.187}$ & $\underset{\left(-0.10\%\right)}{4.185}$ &  & $\underset{\left(2.50\%\right)}{0.164}$ & $\underset{\left(2.48\%\right)}{0.165}$\tabularnewline
$80$ &  & $\underset{\left(-0.07\%\right)}{4.255}$ & $\underset{\left(-0.07\%\right)}{4.097}$ & $\underset{\left(-0.10\%\right)}{4.095}$ &  & $\underset{\left(1.26\%\right)}{0.159}$ & $\underset{\left(1.27\%\right)}{0.160}$\tabularnewline
\midrule 
\multicolumn{8}{l}{GPR-EI}\tabularnewline
$2$ &  & $\underset{\left(-0.09\%\right)}{6.895}$ & $\underset{\left(-0.12\%\right)}{6.651}$ & $\underset{\left(-0.12\%\right)}{6.649}$ &  & $\underset{\left(0.83\%\right)}{0.244}$ & $\underset{\left(0.82\%\right)}{0.246}$\tabularnewline
$10$ &  & $\underset{\left(-0.04\%\right)}{4.864}$ & $\underset{\left(-0.09\%\right)}{4.685}$ & $\underset{\left(-0.09\%\right)}{4.684}$ &  & $\underset{\left(1.13\%\right)}{0.179}$ & $\underset{\left(1.12\%\right)}{0.180}$\tabularnewline
$20$ &  & $\underset{\left(0.00\%\right)}{4.530}$ & $\underset{\left(-0.05\%\right)}{4.362}$ & $\underset{\left(-0.07\%\right)}{4.360}$ &  & $\underset{\left(1.20\%\right)}{0.168}$ & $\underset{\left(1.80\%\right)}{0.170}$\tabularnewline
$40$ &  & $\underset{\left(-0.02\%\right)}{4.349}$ & $\underset{\left(-0.10\%\right)}{4.186}$ & $\underset{\left(-0.12\%\right)}{4.184}$ &  & $\underset{\left(2.50\%\right)}{0.164}$ & $\underset{\left(2.48\%\right)}{0.165}$\tabularnewline
$80$ &  & $\underset{\left(-0.07\%\right)}{4.254}$ & $\underset{\left(0.00\%\right)}{4.100}$ & $\underset{\left(0.05\%\right)}{4.097}$ &  & $\underset{\left(-1.91\%\right)}{0.154}$ & $\underset{\left(-1.26\%\right)}{0.156}$\tabularnewline
\bottomrule
\end{tabular}
\par\end{centering}
\caption{\label{tab:PUT_GEO_1}Numerical results for a Geometric American put
option. Values in brackets are the relative errors with respect to
the American benchmark. $d$ stands for the dimension.}
\end{table}

Table \ref{tab:PUT_GEO_1} shows the numerical results, which appear
to be very accurate and reliable. As far as the price calculation
is considered, the relative errors compared to the American benchmark
never exceed (in absolute value) $0.14\%$, which is a very small
value. The results are even more interesting when compared to the
Bermudian benchmark: in this case, the relative error is always below
$0.07\%$. We can therefore say that, in general, the Bermudian approximation
and the algorithmic approximations have a similar contribution to
the total error with respect to the American price. The relative error
with respect to the XVA are generally larger because the XVA is obtained
as the difference of almost equal quantities, so the absolute error
must be related to a smaller quantity. However, for the cases considered,
the absolute error on the XVA never exceeds $2.50\%$ and, in general,
tends to increase as the problem size increases. Again, the Bermudian
approximation contributes about half of the total error. To conclude,
we observe that the results for $M=V$ and $M=\hat{V}$ are very similar,
both in terms of prices and XVA. 

To investigate the convergence rate of the two methods, we compute
the XVA by changing the number $P$ of points employed for the sparse
quasi-random grid. As one may observe from the results reported in
Table \ref{tab:PUT_GEO_2}, the GPR algorithms provide convergence
to Bermudian prices with great accuracy. Moreover, due to the use
of the control variate technique, very few points are needed to obtain
very accurate results. Obviously, the larger the dimension, the more
points are required to approach the exact value. This fact is particularly
important as the computational time increases more than linearly as
the number of points increases (the higher cost is due to the training
of the GPR model, which is cubic). Finally, we note that the GPR-EI
method is generally faster and more accurate than GPR-MC, but the
latter returns more accurate results in very high dimensions, especially
for $d=80$. 

\begin{table}
\begin{centering}
\scalebox{0.90}{%
\begin{tabular}{cccccccccccccccc}
\toprule 
 &  & \multicolumn{2}{c}{Benchmarks} &  & \multicolumn{5}{c}{GPR-MC} &  &  & \multicolumn{4}{c}{GPR-EI}\tabularnewline
\midrule 
 &  &  &  &  & \multicolumn{11}{c}{$P$}\tabularnewline
$d$ &  & {\footnotesize{}American} & {\footnotesize{}Bermudian} &  & $125$ & $\phantom{1}250$ & $\phantom{1}500$ & $1000$ & $2000$ &  & $\phantom{0}125$ & $\phantom{1}250$ & $\phantom{1}500$ & $1000$ & $2000$\tabularnewline
\midrule
\multicolumn{4}{l}{XVA, case $M=V$} &  &  &  &  &  &  & \tabularnewline
$2$ &  & $0.242$ & $0.244$ &  & $\underset{\left(117\right)}{0.247}$ & $\underset{\left(202\right)}{0.245}$ & $\underset{\left(480\right)}{0.244}$ & $\underset{\left(1419\right)}{0.245}$ & $\underset{\left(4732\right)}{0.244}$ &  & $\underset{\left(103\right)}{0.242}$ & $\underset{\left(126\right)}{0.244}$ & $\underset{\left(151\right)}{0.243}$ & $\underset{\left(320\right)}{0.245}$ & $\underset{\left(841\right)}{0.244}$\tabularnewline
$10$ &  & $0.177$ & $0.178$ &  & $\underset{\left(160\right)}{0.183}$ & $\underset{\left(287\right)}{0.180}$ & $\underset{\left(688\right)}{0.180}$ & $\underset{\left(1784\right)}{0.179}$ & $\underset{\left(5621\right)}{0.179}$ &  & $\underset{\left(132\right)}{0.180}$ & $\underset{\left(199\right)}{0.179}$ & $\underset{\left(358\right)}{0.179}$ & $\underset{\left(698\right)}{0.179}$ & $\underset{\left(1803\right)}{0.179}$\tabularnewline
$20$ &  & $0.166$ & $0.167$ &  & $\underset{\left(243\right)}{0.171}$ & $\underset{\left(447\right)}{0.171}$ & $\underset{\left(1052\right)}{0.172}$ & $\underset{\left(2345\right)}{0.170}$ & $\underset{\left(6592\right)}{0.169}$ &  & $\underset{\left(221\right)}{0.169}$ & $\underset{\left(344\right)}{0.169}$ & $\underset{\left(692\right)}{0.169}$ & $\underset{\left(1257\right)}{0.168}$ & $\underset{\left(2679\right)}{0.168}$\tabularnewline
$40$ &  & $0.160$ & $0.161$ &  & $\underset{\left(363\right)}{0.152}$ & $\underset{\left(614\right)}{0.159}$ & $\underset{\left(1413\right)}{0.164}$ & $\underset{\left(3360\right)}{0.165}$ & $\underset{\left(8537\right)}{0.164}$ &  & $\underset{\left(360\right)}{0.159}$ & $\underset{\left(515\right)}{0.163}$ & $\underset{\left(1076\right)}{0.164}$ & $\underset{\left(2171\right)}{0.164}$ & $\underset{\left(4284\right)}{0.164}$\tabularnewline
$80$ &  & $0.157$ & $0.158$ &  & $\underset{\left(477\right)}{0.118}$ & $\underset{\left(1047\right)}{0.138}$ & $\underset{\left(2287\right)}{0.155}$ & $\underset{\left(4709\right)}{0.158}$ & $\underset{\left(12258\right)}{0.159}$ &  & $\underset{\left(441\right)}{0.070}$ & $\underset{\left(1010\right)}{0.119}$ & $\underset{\left(1885\right)}{0.145}$ & $\underset{\left(3961\right)}{0.151}$ & $\underset{\left(7242\right)}{0.154}$\tabularnewline
\midrule
\multicolumn{6}{l}{XVA, case $M=\hat{V}$} &  &  &  &  & \tabularnewline
$2$ &  & $0.244$ & $0.246$ &  & $\underset{\left(125\right)}{0.249}$ & $\underset{\left(194\right)}{0.247}$ & $\underset{\left(518\right)}{0.249}$ & $\underset{\left(1472\right)}{0.248}$ & $\underset{\left(4869\right)}{0.246}$ &  & $\underset{\left(97\right)}{0.244}$ & $\underset{\left(236\right)}{0.246}$ & $\underset{\left(301\right)}{0.245}$ & $\underset{\left(607\right)}{0.247}$ & $\underset{\left(1570\right)}{0.246}$\tabularnewline
$10$ &  & $0.178$ & $0.180$ &  & $\underset{\left(190\right)}{0.188}$ & $\underset{\left(301\right)}{0.183}$ & $\underset{\left(702\right)}{0.181}$ & $\underset{\left(1811\right)}{0.182}$ & $\underset{\left(5528\right)}{0.181}$ &  & $\underset{\left(132\right)}{0.182}$ & $\underset{\left(367\right)}{0.181}$ & $\underset{\left(680\right)}{0.181}$ & $\underset{\left(1286\right)}{0.181}$ & $\underset{\left(3252\right)}{0.181}$\tabularnewline
$20$ &  & $0.167$ & $0.169$ &  & $\underset{\left(249\right)}{0.175}$ & $\underset{\left(464\right)}{0.175}$ & $\underset{\left(1011\right)}{0.174}$ & $\underset{\left(2295\right)}{0.172}$ & $\underset{\left(6680\right)}{0.171}$ &  & $\underset{\left(212\right)}{0.170}$ & $\underset{\left(600\right)}{0.170}$ & $\underset{\left(1276\right)}{0.170}$ & $\underset{\left(2275\right)}{0.170}$ & $\underset{\left(4830\right)}{0.170}$\tabularnewline
$40$ &  & $0.161$ & $0.163$ &  & $\underset{\left(353\right)}{0.152}$ & $\underset{\left(657\right)}{0.163}$ & $\underset{\left(1345\right)}{0.165}$ & $\underset{\left(3365\right)}{0.167}$ & $\underset{\left(8782\right)}{0.165}$ &  & $\underset{\left(355\right)}{0.160}$ & $\underset{\left(954\right)}{0.164}$ & $\underset{\left(1982\right)}{0.166}$ & $\underset{\left(3997\right)}{0.165}$ & $\underset{\left(7853\right)}{0.165}$\tabularnewline
$80$ &  & $0.158$ & $0.159$ &  & $\underset{\left(503\right)}{0.119}$ & $\underset{\left(1160\right)}{0.141}$ & $\underset{\left(2242\right)}{0.156}$ & $\underset{\left(5020\right)}{0.159}$ & $\underset{\left(12265\right)}{0.160}$ &  & $\underset{\left(832\right)}{0.071}$ & $\underset{\left(1804\right)}{0.120}$ & $\underset{\left(3366\right)}{0.147}$ & $\underset{\left(6826\right)}{0.153}$ & $\underset{\left(7395\right)}{0.156}$\tabularnewline
\bottomrule
\end{tabular}}
\par\end{centering}
\caption{\label{tab:PUT_GEO_2}Numerical results for a Geometric American put
option. Values in brackets are the computational times (in seconds).
$P$ is the number of points employed in the GPR algorithms.}
\end{table}

\subsection*{Call on the maximum}

The American option Call on the maximum is a difficult to evaluate
derivative and so it has been considered by many authors, such as
Schoenmakers \cite{schoenmakers2013optimal}, Lelong \cite{lelong2018dual},
Becker et al. \cite{becker2019deep}, Goudenège et al. \cite{goudenege2020machine,goudenege2021variance},
and Ech-Chafiq et al. \cite{ech2021pricing}. Specifically, the payoff
of such an option is given by
\[
H(\mathbf{S}_{T})=\left(\max_{i=1,\dots,d}S_{T}^{i}-K\right)_{+}.
\]

We start the numerical analysis by considering the same model parameters
as for the Geometric put, which are reported in Table \ref{tab:BS}.
We stress out that, since the considered derivative is a call option
and the underlying pays no dividends ($\eta_{i}=0$, see Table \ref{tab:BS}),
early exercise is never optimal for the riskless option. Moreover,
since the payoff of the derivative is always possible, we can use
the closed formulas proposed by Burgard and Kjaer \cite{burgard2010pde}
to compute the XVA for the European derivative. Specifically, if $M=V$,
then 
\[
XVA^{EU}=V^{EU}\left(t,\mathbf{S}_{0}\right)\cdot\left(1-e^{-\left(\lambda_{B}+\lambda_{C}\right)T}-c_{p}\frac{1-e^{-\left(\lambda_{B}+\lambda_{C}\right)T}}{\lambda_{B}+\lambda_{C}}\right),
\]
and if $M=\hat{V},$ then
\[
XVA^{EU}=V^{EU}\left(t,\mathbf{S}_{0}\right)\cdot\left(1-e^{\left(c_{p}-\lambda_{B}-\lambda_{C}\right)T}\right).
\]

It is worth noting that despite the prices of an European and an American
riskless options are equals, this does not also apply to their XVAs.
In fact, an American option may be exercised early so to reduce the
losses due to counterparty default, thus the XVA on the American option
is expected to be smaller than the European one. So, we present the
XVA on the European option as an upper-bound (UB). Results are shown
in Table \ref{tab:CALL_MAX_1}. We can see that both proposed methods
provide very accurate values for the cases considered. When a large
number of points is used (at least 500), the relative deviation between
the returned values, that is the difference divided by the larger
value, is less than 2\%. The values obtained for XVA are all below
the upper-bound, although very close to it.

\begin{table}
\begin{centering}
\scalebox{0.95}{%
\begin{tabular}{ccccccccccccccc}
\toprule 
$d$ &  & \multicolumn{5}{c}{GPR-MC} &  & \multicolumn{5}{c}{GPR-EI} &  & UB\tabularnewline
\midrule 
 &  & \multicolumn{11}{c}{$P$} &  & \tabularnewline
\cmidrule{3-13} \cmidrule{4-13} \cmidrule{5-13} \cmidrule{6-13} \cmidrule{7-13} \cmidrule{8-13} \cmidrule{9-13} \cmidrule{10-13} \cmidrule{11-13} \cmidrule{12-13} \cmidrule{13-13} 
 &  & $\phantom{0}125$ & $\phantom{1}250$ & $\phantom{1}500$ & $1000$ & $2000$ &  & $\phantom{0}125$ & $\phantom{1}250$ & $\phantom{1}500$ & $1000$ & $2000$ &  & \tabularnewline
\multicolumn{7}{l}{XVA, case $M=V$} &  &  &  &  & \tabularnewline
$2$ &  & $1.004$ & $1.001$ & $0.998$ & $0.999$ & $0.998$ &  & $0.999$ & $0.999$ & $0.999$ & $0.998$ & $0.998$ &  & $1.009\pm0.001$\tabularnewline
$10$ &  & $2.191$ & $2.210$ & $2.226$ & $2.231$ & $2.229$ &  & $2.237$ & $2.231$ & $2.236$ & $2.234$ & $2.230$ &  & $2.252\pm0.001$\tabularnewline
$20$ &  & $2.670$ & $2.720$ & $2.753$ & $2.762$ & $2.771$ &  & $2.789$ & $2.792$ & $2.788$ & $2.774$ & $2.774$ &  & $2.803\pm0.001$\tabularnewline
$40$ &  & $2.743$ & $3.149$ & $3.223$ & $3.268$ & $3.294$ &  & $3.317$ & $3.262$ & $3.247$ & $3.296$ & $3.299$ &  & $3.337\pm0.001$\tabularnewline
$80$ &  & $2.674$ & $3.001$ & $3.602$ & $3.702$ & $3.764$ &  & $2.830$ & $3.241$ & $3.642$ & $3.736$ & $3.741$ &  & $3.852\pm0.001$\tabularnewline
\midrule
\multicolumn{7}{l}{XVA, case $M=\hat{V}$} &  &  &  &  & \tabularnewline
$2$ &  & $1.012$ & $1.019$ & $1.011$ & $1.010$ & $1.011$ &  & $1.010$ & $1.011$ & $1.011$ & $1.010$ & $1.011$ &  & $1.021\pm0.001$\tabularnewline
$10$ &  & $2.217$ & $2.238$ & $2.250$ & $2.254$ & $2.255$ &  & $2.263$ & $2.257$ & $2.261$ & $2.259$ & $2.257$ &  & $2.279\pm0.001$\tabularnewline
$20$ &  & $2.711$ & $2.748$ & $2.787$ & $2.790$ & $2.803$ &  & $2.820$ & $2.825$ & $2.823$ & $2.806$ & $2.809$ &  & $2.837\pm0.001$\tabularnewline
$40$ &  & $2.766$ & $3.180$ & $3.259$ & $3.305$ & $3.333$ &  & $3.358$ & $3.298$ & $3.284$ & $3.336$ & $3.339$ &  & $3.377\pm0.001$\tabularnewline
$80$ &  & $2.705$ & $3.046$ & $3.646$ & $3.742$ & $3.812$ &  & $2.891$ & $3.259$ & $3.699$ & $3.732$ & $3.783$ &  & $3.898\pm0.001$\tabularnewline
\bottomrule
\end{tabular}}
\par\end{centering}
\caption{\label{tab:CALL_MAX_1}Numerical results for a Call on the maximum
option. The confidence interval for the upper-bound UB is computed
at a $99\%$ confidence level.}
\end{table}

Finally, for the sake of comparison, let us calculate the XVA for
a Call on the maximum considering a positive dividend rate, equal
for all underlyings and equal to $\eta_{i}=0.02$. In this specific
case, there are neither benchmarks nor upper-bounds. Table \ref{tab:CALL_MAX_2}
presents the results. In the case under consideration (positive dividend),
the valuation seems to be more challenging than in the previous case
(with zero dividend). In fact, at least 500 points are needed to obtain
a relative deviation of less than 5\%.

\begin{table}
\begin{centering}
\scalebox{0.95}{%
\begin{tabular}{ccccccccccccc}
\toprule 
$d$ &  & \multicolumn{5}{c}{GPR-MC} &  & \multicolumn{5}{c}{GPR-EI}\tabularnewline
\midrule 
 &  & \multicolumn{11}{c}{$P$}\tabularnewline
\cmidrule{3-13} \cmidrule{4-13} \cmidrule{5-13} \cmidrule{6-13} \cmidrule{7-13} \cmidrule{8-13} \cmidrule{9-13} \cmidrule{10-13} \cmidrule{11-13} \cmidrule{12-13} \cmidrule{13-13} 
 &  & $\phantom{0}125$ & $\phantom{1}250$ & $\phantom{1}500$ & $1000$ & $2000$ &  & $\phantom{0}125$ & $\phantom{1}250$ & $\phantom{1}500$ & $1000$ & $2000$\tabularnewline
\multicolumn{7}{l}{XVA, case $M=V$} &  &  & \tabularnewline
$2$ &  & $0.817$ & $0.805$ & $0.800$ & $0.806$ & $0.804$ &  & $0.805$ & $0.801$ & $0.802$ & $0.802$ & $0.803$\tabularnewline
$10$ &  & $1.988$ & $2.003$ & $2.008$ & $2.009$ & $2.006$ &  & $1.962$ & $1.966$ & $2.000$ & $2.004$ & $1.999$\tabularnewline
$20$ &  & $2.479$ & $2.529$ & $2.551$ & $2.555$ & $2.556$ &  & $2.473$ & $2.485$ & $2.481$ & $2.456$ & $2.457$\tabularnewline
$40$ &  & $2.599$ & $2.948$ & $3.009$ & $3.063$ & $3.080$ &  & $3.026$ & $2.973$ & $2.964$ & $3.005$ & $2.997$\tabularnewline
$80$ &  & $2.577$ & $2.850$ & $3.407$ & $3.487$ & $3.557$ &  & $2.811$ & $3.033$ & $3.461$ & $3.422$ & $3.488$\tabularnewline
\midrule
\multicolumn{7}{l}{XVA, case $M=\hat{V}$} &  &  & \tabularnewline
$2$ &  & $0.820$ & $0.814$ & $0.809$ & $0.808$ & $0.815$ &  & $0.814$ & $0.811$ & $0.812$ & $0.811$ & $0.812$\tabularnewline
$10$ &  & $2.018$ & $2.031$ & $2.027$ & $2.030$ & $2.031$ &  & $1.981$ & $1.992$ & $2.025$ & $2.026$ & $2.022$\tabularnewline
$20$ &  & $2.505$ & $2.555$ & $2.581$ & $2.581$ & $2.588$ &  & $2.503$ & $2.514$ & $2.508$ & $2.488$ & $2.484$\tabularnewline
$40$ &  & $2.625$ & $2.978$ & $3.051$ & $3.100$ & $3.118$ &  & $3.059$ & $3.009$ & $2.998$ & $3.040$ & $3.033$\tabularnewline
$80$ &  & $2.613$ & $2.878$ & $3.449$ & $3.531$ & $3.597$ &  & $2.848$ & $3.068$ & $3.461$ & $3.464$ & $3.560$\tabularnewline
\bottomrule
\end{tabular}}
\par\end{centering}
\caption{\label{tab:CALL_MAX_2}Numerical results for a Call on the maximum
option with a positive dividend rate. }
\end{table}

\subsection*{Swaption with floor on two portfolios}

The derivatives considered in the numerical examples above are all
options and therefore their payoff function and their values are always
positive. The model considered in this work also admits negative values
for the payout, so it is interesting to consider a case with this
attribute. Let us now consider an American two-portfolio Swaption
with a negative floor, i.e. a derivative in which two portfolios are
swapped between counterparties, whose value can be either positive
or negative. Specifically, the first portfolio consists of the first
$d/2$ underlyings and the second portfolio consists of the remaining
underlyings. For simplicity, we will assume $d$ to be an even number.
In both cases, the underlyings all have the same weight, so the value
of each portfolio is equal to the average of the prices of the individual
risky assets. The payout of such a derivative is given by
\[
H(\mathbf{S}_{T})=\max\left(\frac{2}{d}\left(\sum_{i=1}^{d/2}S_{T}^{i}-\sum_{i=d/2+1}^{d}S_{T}^{i}\right),K\right).
\]
 In particular, the floor $K$ is a negative number, thus the payout
of the option can be negative. Table \ref{tab:PTF_SWAP} presents
the numerical results. We observe that, in the case considered, the
estimated values for the XVA are much smaller than in the previous
cases. The two methods return very similar values for $d\leq40$,
whereas for $d=80$ GPR-EI estimates of the XVA are greater than those
returned by GPR-MC (approximately $+20\%$). The lack of a benchmark
makes it unclear which of the two methods is the more accurate in
this case.

\begin{table}
\begin{centering}
\scalebox{0.95}{%
\begin{tabular}{ccrrrrrrccccc}
\toprule 
$d$ &  & \multicolumn{5}{c}{GPR-MC} &  & \multicolumn{5}{c}{GPR-EI}\tabularnewline
\midrule 
 &  & \multicolumn{11}{c}{$P$}\tabularnewline
\cmidrule{3-13} \cmidrule{4-13} \cmidrule{5-13} \cmidrule{6-13} \cmidrule{7-13} \cmidrule{8-13} \cmidrule{9-13} \cmidrule{10-13} \cmidrule{11-13} \cmidrule{12-13} \cmidrule{13-13} 
 &  & $\phantom{0}125$ & $\phantom{1}250$ & $\phantom{1}500$ & $1000$ & $2000$ &  & $\phantom{0}125$ & $\phantom{1}250$ & $\phantom{1}500$ & $1000$ & $2000$\tabularnewline
\multicolumn{7}{l}{XVA, case $M=V$} &  &  & \tabularnewline
$2$ &  & $41.911$ & $42.031$ & $41.953$ & $41.884$ & $42.054$ &  & $42.086$ & $41.958$ & $42.063$ & $41.982$ & $42.000$\tabularnewline
$10$ &  & $14.604$ & $14.938$ & $14.693$ & $14.762$ & $14.750$ &  & $14.820$ & $14.307$ & $14.551$ & $14.691$ & $14.739$\tabularnewline
$20$ &  & $\phantom{1}7.959$ & $\phantom{1}8.168$ & $\phantom{1}8.567$ & $\phantom{1}8.631$ & $\phantom{1}8.827$ &  & $\phantom{1}7.612$ & $\phantom{1}7.560$ & $\phantom{1}8.241$ & $\phantom{1}8.230$ & $\phantom{1}8.441$\tabularnewline
$40$ &  & $\phantom{1}4.426$ & $\phantom{1}4.546$ & $\phantom{1}4.277$ & $\phantom{1}4.397$ & $\phantom{1}4.427$ &  & $\phantom{1}3.710$ & $\phantom{1}4.010$ & $\phantom{1}4.275$ & $\phantom{1}4.215$ & $\phantom{1}4.354$\tabularnewline
$80$ &  & $\phantom{1}3.010$ & $\phantom{1}1.448$ & $\phantom{1}1.855$ & $\phantom{1}2.079$ & $\phantom{1}2.089$ &  & $\phantom{1}2.173$ & $\phantom{1}2.287$ & $\phantom{1}2.613$ & $\phantom{1}2.524$ & $\phantom{1}2.474$\tabularnewline
\midrule
\multicolumn{7}{l}{XVA, case $M=\hat{V}$} &  &  & \tabularnewline
$2$ &  & $42.560$ & $42.404$ & $42.372$ & $42.563$ & $42.463$ &  & $42.594$ & $42.377$ & $42.477$ & $42.377$ & $42.549$\tabularnewline
$10$ &  & $14.894$ & $15.040$ & $14.789$ & $14.849$ & $14.897$ &  & $14.941$ & $14.369$ & $14.694$ & $14.775$ & $14.854$\tabularnewline
$20$ &  & $\phantom{1}8.022$ & $\phantom{1}8.233$ & $\phantom{1}8.631$ & $\phantom{1}8.653$ & $\phantom{1}8.937$ &  & $\phantom{1}7.587$ & $\phantom{1}7.726$ & $\phantom{1}8.355$ & $\phantom{1}8.383$ & $\phantom{1}8.592$\tabularnewline
$40$ &  & $\phantom{1}4.446$ & $\phantom{1}4.589$ & $\phantom{1}4.282$ & $\phantom{1}4.414$ & $\phantom{1}4.559$ &  & $\phantom{1}3.681$ & $\phantom{1}4.081$ & $\phantom{1}4.265$ & $\phantom{1}4.249$ & $\phantom{1}4.412$\tabularnewline
$80$ &  & $\phantom{1}3.179$ & $\phantom{1}2.493$ & $\phantom{1}1.899$ & $\phantom{1}2.104$ & $\phantom{1}2.132$ &  & $\phantom{1}2.146$ & $\phantom{1}2.301$ & $\phantom{1}2.682$ & $\phantom{1}2.611$ & $\phantom{1}2.593$\tabularnewline
\bottomrule
\end{tabular}}
\par\end{centering}
\caption{\label{tab:PTF_SWAP}Numerical results for a Swaption with floor on
two portfolios. All the results must be multiplied by $10^{-2}$.}
\end{table}

\section{Conclusion}

In this paper, we have discussed the problem of calculating the XVA
of a derivative that depends on multiple underlyings. This issue plays
an essential role in counterparty risk management, also in light of
the regulations currently in force. Nevertheless, it is an element
that is often overlooked due to the curse of dimensionality associated
with the problem of valuing high-dimensional options. Our proposal
to address this challenge is to reformulate the problem in probabilistic
terms and make use of the GPR-MC and GPR-EI techniques with control
variate, which have already been successfully applied in similar contexts.
Numerical results show that it is possible to obtain very accurate
estimates of the XVA and, in some cases, very few points are sufficient
to achieve very accurate results. For the considered cases, the proposed
methods demonstrate to be effective for large dimensions, thus providing
new methods for estimating XVA by overcoming the curse of dimensionality.

\bibliographystyle{abbrv}
\bibliography{bibliography}

\appendix

\section{\label{App1}Proof of Proposition \ref{Prop1} }

Equation (\ref{eq:implicitVtilde}) is a non-linear equation, so,
first, we discuss existence and uniqueness of the solution. Let us
assume that we have fixed the value of $t_{n}$ and $\mathbf{x}$,
so we can consider them as model parameters. We define the function
$f:\mathbb{R}\rightarrow\mathbb{R}$ as
\[
f_{t_{n},\mathbf{x}}\left(z\right)=\max\left\{ E\left(\mathbf{x}\right)+\frac{\Delta t}{2}\left(z^{+}c_{p}+z^{-}c_{m}\right),H\left(\mathbf{x}\right)\right\} ,
\]
so that equation (\ref{eq:implicitVtilde}) can be rewritten as 
\[
\tilde{V}\left(t_{n,},\mathbf{x}\right)=f_{t_{n},\mathbf{x}}\left(\tilde{V}\left(t_{n,},\mathbf{x}\right)\right).
\]
So, in to $\tilde{V}\left(t_{n,},\mathbf{x}\right)$ one has to solve
the equation $z-f_{t_{n},\mathbf{x}}\left(z\right)=0$, that is computing
the zeros of the function $F_{t_{n},\mathbf{x}}\left(z\right)=x-f_{t_{n},\mathbf{x}}\left(z\right)$.
We observe that $F_{t_{n},\mathbf{x}}$ is a continuous function and
it is piecewise derivable. In particular, if 
\[
z\neq0,\quad z\neq2\frac{H\left(\mathbf{x}\right)-E\left(\mathbf{x}\right)}{c_{p}\Delta t},\ \text{and}\ z\neq2\frac{H\left(\mathbf{x}\right)-E\left(\mathbf{x}\right)}{c_{m}\Delta t},
\]
 then the derivative of $F_{t_{n},\mathbf{x}}$ is given by
\[
\frac{d}{dz}\left(F_{t_{n},\mathbf{x}}\left(z\right)\right)=\begin{cases}
1-\frac{\Delta t}{2}c_{p} & \text{if }z\geq0\text{ and }E\left(\mathbf{x}\right)+\frac{\Delta t}{2}c_{p}z>H\left(\mathbf{x}\right),\\
1-\frac{\Delta t}{2}c_{m} & \text{if }z<0\text{ and }E\left(\mathbf{x}\right)+\frac{\Delta t}{2}c_{m}z>H\left(\mathbf{x}\right),\\
1 & \text{if }E\left(\mathbf{x}\right)+\frac{\Delta t}{2}\left(z^{+}c_{p}+z^{-}c_{m}\right)<H\left(\mathbf{x}\right),
\end{cases}
\]
that is
\[
\frac{d}{dz}\left(F_{t_{n},\mathbf{x}}\left(z\right)\right)=\begin{cases}
1-\frac{\Delta t}{2}c_{p} & \text{if }z>0\text{ and }z>2\frac{H\left(\mathbf{x}\right)-E\left(\mathbf{x}\right)}{\Delta tc_{p}},\\
1-\frac{\Delta t}{2}c_{m} & \text{if }z<0\text{ and }z>2\frac{H\left(\mathbf{x}\right)-E\left(\mathbf{x}\right)}{\Delta tc_{m}},\\
1 & \text{otherwise}.
\end{cases}
\]
Therefore $F_{t_{n},\mathbf{x}}$ is a continuous piecewise linear
function. Moreover, if we assume $1-\frac{\Delta t}{2}c_{p}>0$ and
$1-\frac{\Delta t}{2}c_{m}>0$ (which is true for $\Delta t$ small
enough) $F_{t_{n},\mathbf{x}}$ is strictly increasing, so it can
not have more than one zero. Furthermore, we observe

\[
\lim_{z\rightarrow-\infty}F_{t_{n},\mathbf{x}}\left(z\right)=-\infty,\quad\lim_{z\rightarrow+\infty}F_{t_{n},\mathbf{x}}\left(z\right)=+\infty,
\]
 so there is one and only one solution to $F_{t_{n},\mathbf{x}}\left(z\right)=0$.

Now, we have proved that there is one and only one solution, let us
compute it. We rewrite equation (\ref{eq:implicitVtilde}) as 

\[
\tilde{V}\left(t_{n,},\mathbf{x}\right)=\max\left\{ E\left(\mathbf{x}\right)+\frac{\Delta t}{2}\left(\tilde{V}\left(t_{n,},\mathbf{x}\right)^{+}c_{p}+\tilde{V}\left(t_{n,},\mathbf{x}\right)^{-}c_{m}\right),H\left(\mathbf{x}\right)\right\} .
\]
 We distinguish 5 cases.

\paragraph*{Case 1a: $\tilde{V}\left(t_{n,},\mathbf{x}\right)=H\left(\mathbf{x}\right)\protect\leq0$.}

In this case, we have 
\[
\max\left\{ E\left(\mathbf{x}\right)+\frac{\Delta t}{2}\tilde{V}\left(t_{n,},\mathbf{x}\right),H\left(\mathbf{x}\right)\right\} =H\left(\mathbf{x}\right),
\]
so 
\[
E\left(\mathbf{x}\right)+\frac{\Delta t}{2}H\left(\mathbf{x}\right)c_{m}=E\left(\mathbf{x}\right)+\frac{\Delta t}{2}\tilde{V}\left(t_{n,},\mathbf{x}\right)c_{m}\leq H\left(\mathbf{x}\right),
\]
thus
\[
E\left(\mathbf{x}\right)\leq H\left(\mathbf{x}\right)\left(1-\frac{\Delta t}{2}c_{m}\right)\leq0.
\]

\paragraph*{Case 1b: $H\left(\mathbf{x}\right)<\tilde{V}\left(t_{n,},\mathbf{x}\right)\protect\leq0$.}

In this case, we have 
\[
\max\left\{ E\left(\mathbf{x}\right)+\frac{\Delta t}{2}\tilde{V}\left(t_{n,},\mathbf{x}\right)c_{m},H\left(\mathbf{x}\right)\right\} =E\left(\mathbf{x}\right)+\frac{\Delta t}{2}\tilde{V}\left(t_{n,},\mathbf{x}\right)c_{m}=\tilde{V}\left(t_{n,},\mathbf{x}\right),
\]
so 
\[
\tilde{V}\left(t_{n,},\mathbf{x}\right)=\frac{E\left(\mathbf{x}\right)}{1-\frac{\Delta t}{2}c_{m}},
\]
which implies $E\leq0$ and
\[
H\left(\mathbf{x}\right)\left(1-\frac{\Delta t}{2}c_{m}\right)<\tilde{V}\left(t_{n,},\mathbf{x}\right)\left(1-\frac{\Delta t}{2}c_{m}\right)=E\left(\mathbf{x}\right)\leq0.
\]

\paragraph*{Case 1c: $H\left(\mathbf{x}\right)<0<\tilde{V}\left(t_{n,},\mathbf{x}\right)$.}

In this case, we have 
\[
\max\left\{ E\left(\mathbf{x}\right)+\frac{\Delta t}{2}\tilde{V}\left(t_{n,},\mathbf{x}\right)c_{p},H\left(\mathbf{x}\right)\right\} =E\left(\mathbf{x}\right)+\frac{\Delta t}{2}\tilde{V}\left(t_{n,},\mathbf{x}\right)c_{p}=\tilde{V}\left(t_{n,},\mathbf{x}\right),
\]
so 
\[
\tilde{V}\left(t_{n,},\mathbf{x}\right)=\frac{E\left(\mathbf{x}\right)}{1-\frac{\Delta t}{2}c_{p}},
\]
and, since $E\left(\mathbf{x}\right)\geq0,$we also have
\[
H\left(\mathbf{x}\right)\left(1-\frac{\Delta t}{2}c_{m}\right)<0<E\left(\mathbf{x}\right).
\]

\paragraph*{Case 2a: $0<\tilde{V}\left(t_{n,},\mathbf{x}\right)=H\left(\mathbf{x}\right)$.}

In this case, we have 
\[
\max\left\{ E\left(\mathbf{x}\right)+\frac{\Delta t}{2}\tilde{V}\left(t_{n,},\mathbf{x}\right)c_{p},H\left(\mathbf{x}\right)\right\} =H\left(\mathbf{x}\right),
\]
so 
\[
E\left(\mathbf{x}\right)+\frac{\Delta t}{2}H\left(\mathbf{x}\right)c_{p}=E\left(\mathbf{x}\right)+\frac{\Delta t}{2}\tilde{V}\left(t_{n,},\mathbf{x}\right)c_{p}\leq H\left(\mathbf{x}\right),
\]
thus
\[
E\left(\mathbf{x}\right)\leq H\left(\mathbf{x}\right)\left(1-\frac{\Delta t}{2}c_{p}\right).
\]

\paragraph*{Case 2b: $0\protect\leq H\left(\mathbf{x}\right)<\hat{V}(t_{n,},S)$.}

In this case, we have 
\[
\max\left\{ E\left(\mathbf{x}\right)+\frac{\Delta t}{2}\tilde{V}\left(t_{n,},\mathbf{x}\right)c_{p},H\left(\mathbf{x}\right)\right\} =E\left(\mathbf{x}\right)+\frac{\Delta t}{2}\tilde{V}\left(t_{n,},\mathbf{x}\right)c_{p}=\tilde{V}\left(t_{n,},\mathbf{x}\right),
\]
so 
\[
\tilde{V}\left(t_{n,},\mathbf{x}\right)=\frac{E\left(\mathbf{x}\right)}{1-\frac{\Delta t}{2}c_{p}},
\]
thus 
\[
E\left(\mathbf{x}\right)>H\left(\mathbf{x}\right)\left(1-\frac{\Delta t}{2}c_{p}\right).
\]

\vspace{3mm}

So, cases 1a, 1b, 1c, 2a, 2b, which define a partition of the possible,
induce 5 possible relations between $E\left(\mathbf{x}\right)$ and
$H\left(\mathbf{x}\right)$ which are incompatible and exhaustive.
Let us summarize these relations:
\begin{enumerate}
\item If $H\left(\mathbf{x}\right)\leq0$ and $E\left(\mathbf{x}\right)\leq H\left(\mathbf{x}\right)\left(1-\frac{\Delta t}{2}c_{m}\right)\leq0$
then case 1a holds and $\tilde{V}\left(t_{n,},\mathbf{x}\right)=H\left(\mathbf{x}\right)$;
\item If $H\left(\mathbf{x}\right)\leq0$ and $H\left(\mathbf{x}\right)\left(1-\frac{\Delta t}{2}c_{m}\right)<E\left(\mathbf{x}\right)\leq0$
then case 1b holds and $\tilde{V}\left(t_{n,},\mathbf{x}\right)=\frac{E\left(\mathbf{x}\right)}{1-\frac{\Delta t}{2}c_{m}}$;
\item If $H\left(\mathbf{x}\right)\leq0$ and $0<E\left(\mathbf{x}\right)$
then case 1c holds and $\tilde{V}\left(t_{n,},\mathbf{x}\right)=\frac{E\left(\mathbf{x}\right)}{1-\frac{\Delta t}{2}c_{p}}$;
\item If $H\left(\mathbf{x}\right)>0$ and $E\left(\mathbf{x}\right)\leq H\left(\mathbf{x}\right)\left(1-\frac{\Delta t}{2}c_{p}\right)$
then case 2a holds and $\tilde{V}\left(t_{n,},\mathbf{x}\right)=H\left(\mathbf{x}\right)$;
\item If $H\left(\mathbf{x}\right)>0$ and $E\left(\mathbf{x}\right)>H\left(\mathbf{x}\right)\left(1-\frac{\Delta t}{2}c_{p}\right)$
then case 2b holds and $\tilde{V}\left(t_{n,},\mathbf{x}\right)=\frac{E\left(\mathbf{x}\right)}{1-\frac{\Delta t}{2}c_{p}}$.
\end{enumerate}
These 5 cases solve the fixed point problem (\ref{eq:implicitVhat}).
\end{document}